\documentclass[journal]{IEEEtran}
\usepackage{subfigure}

\usepackage{stackengine}
\def\delequal{\mathrel{\ensurestackMath{\stackon[1pt]{=}{\scriptstyle\Delta}}}}
\usepackage{hyperref}
\usepackage[all]{hypcap}        
\usepackage{xcolor}
\usepackage[linesnumbered,ruled,vlined]{algorithm2e}
\usepackage{algorithmic}

\SetCommentSty{mycommfont}
\SetKwInput{KwInput}{Input}
\SetKwInput{KwInitialize}{Initialize}
\SetKwInput{KwOutput}{Output} 
\usepackage{xcolor,soul,framed} 
\usepackage{xspace}
\colorlet{shadecolor}{yellow}
\usepackage[pdftex]{graphicx}
\graphicspath{{../pdf/}{../jpeg/}}
\DeclareGraphicsExtensions{.pdf,.jpeg,.png}

\usepackage[cmex10]{amsmath}
\usepackage{array}
\usepackage{mdwmath}
\usepackage{mdwtab}
\usepackage{eqparbox}
\usepackage{url}
\usepackage{cite}

\usepackage{multirow}
\usepackage{multicol}
\usepackage{amsmath,amssymb,amsfonts}
\usepackage{mathtools}
\usepackage{mathrsfs}
\begin{document}
    \title{Design of an Efficient CSI Feedback Mechanism in Massive MIMO Systems: A Machine Learning Approach using Empirical Data}
  \author{M. Karam~Shehzad,~\IEEEmembership{Graduate Student Member,~IEEE,}
      Luca~Rose,~\IEEEmembership{Member,~IEEE,}\\
      Stefan~Wesemann,
      Mohamad~Assaad,~\IEEEmembership{Senior Member,~IEEE,}
      and~Syed Ali~Hassan,~\IEEEmembership{Senior Member,~IEEE}

  \thanks{M. Karam Shehzad is with Nokia Bell-Labs and CentraleSupelec, University of Paris-Saclay, Paris-Saclay, France. Luca Rose is with Nokia Bell-Labs, Paris-Saclay, France. Stefan Wesemann is with Nokia Bell-Labs, Stuttgart, Germany. Mohamad Assaad is with CentraleSupelec, University of Paris-Saclay, Paris-Saclay, France. Syed Ali Hassan is with the School of Electrical Engineering \& Computer Science (SEECS), National University of Sciences \& Technology (NUST), Islamabad, Pakistan. (e-mails: \{muhammad.shehzad, luca.rose, stefan.wesemann\}@nokia.com); mohamad.assaad@centralesupelec.fr; ali.hassan@seecs.edu.pk.}
  \thanks{The corresponding author is M. Karam Shehzad.}

}


\maketitle

\begin{abstract}
Massive multiple-input multiple-output (mMIMO) regime reaps the benefits of spatial diversity and multiplexing gains, subject to precise channel state information (CSI) acquisition. In the current communication architecture, the downlink CSI is estimated by the user equipment (UE) via dedicated pilots and then fed back to the gNodeB (gNB). The feedback information is compressed with the goal of reducing over-the-air overhead. This compression increases the inaccuracy of acquired CSI, thus degrading the overall spectral efficiency. 
This paper proposes a computationally inexpensive machine learning (ML)-based CSI feedback algorithm, which exploits twin channel predictors. The proposed approach can work for both time-division duplex (TDD) and frequency-division duplex (FDD) systems, and it allows to reduce feedback overhead and improves the acquired CSI’s accuracy. To observe real benefits, we demonstrate the performance of the proposed approach using the empirical data recorded at the Nokia campus in Stuttgart, Germany. Numerical results show the effectiveness of the proposed approach in terms of reducing overhead, minimizing quantization errors, increasing spectral efficiency, cosine similarity, and precoding gain compared to the traditional CSI feedback mechanism. 
\end{abstract}


%
\IEEEpeerreviewmaketitle


\section{Introduction}
\IEEEPARstart{O}{ver} the past two decades, multi-antenna techniques dubbed as massive multiple-input multiple-output (mMIMO) have been widely acknowledged as key enablers for the future cellular networks \cite{Intro_mMIMO}. Employing a large number of antennas
in a centralized \cite{centralized} or distributed \cite{distributed} configuration can help in, e.g., reducing multi-user interference \cite{MU_Interference}, multifold increase in channel capacity \cite{mMIMO_capacity}, and cost-effective and reliable coverage by using limited radio resources. To fully harvest mMIMO potential, channel state information (CSI) needs to be acquired at the gNodeB (gNB) side. 

In practical systems, closed loop CSI acquisition, which requires receiver channel estimation over transmitted pilots, is adopted for both time-division duplex (TDD) and frequency-division duplex (FDD) systems.
Consequently, the estimated CSI is sent via feedback links to the gNB. To minimize feedback overhead, codebook\footnote{For the reader's understanding, there are two type of codebooks: type-I and type-II. However, addressing these codebooks is not the focus of this paper, interested readers can refer to, e.g., \cite{Type-I:3GPP}.} (a set of precoding matrix) or quantization-based schemes are adopted, which are prone to compression errors. In mMIMO, feedback overhead grows linearly with the number of antennas. It thus becomes challenging to design an efficient CSI feedback mechanism \cite{lim_fb}. Conversely, in TDD systems, due to the presence of channel reciprocity, CSI at the gNB can be acquired.
Although TDD systems can avoid feedback delays, they are prone to processing delays.  
\subsection{Related Work}
The challenge of CSI feedback in mMIMO systems have been tackled by a few researchers \cite{CSI_FB1, CSI_FB2, CSI_FBVTC, ATG_CS}.  
The authors of \cite{CSI_FB1} and \cite{CSI_FB2} use spatial and temporal correlation of CSI to reduce feedback overhead. This idea has inspired the establishment of protocols for CSI feedback \cite{CSI_FB1, CSI_FB2}. Multi-dimensional compressed sensing (CS) theory, which is based on spatial and frequency correlation of the channel, as well as Tucker decomposition \cite{tucker}, is used in \cite{CSI_FBVTC} to reduce feedback overhead. In addition, an antenna grouping-based feedback reduction method is proposed in \cite{ATG_CS}, where multiple correlated antenna elements are mapped to a single value by exploiting predesigned patterns.   
Similarly, many algorithms have been proposed in CS \cite{CSI_FB3, CSI_FB4}, which cannot fully recover compressed CSI due to the use of simple sparsity prior while the channel matrix used is not perfectly but is \textit{approximately} sparse. All these algorithms do not recover accurate CSI, and a massive amount of data processing is required, making them infeasible.  

The alternative of codebook-based CSI acquisition has been explored in \cite{11, 12, 10, 13, rana2}. This technique is at the base of various commercial systems, such as long-term evolution (LTE)/LTE-advanced and fifth-generation new-radio (5G-NR) that reduce over-the-air feedback. Broadly, the channel representation or precoding matrix is chosen from a shared and known set of matrices, known as codebook.
By considering line-of-sight (LoS) and non-LoS (NLoS) components between the gNB and user equipment (UE), codebooks are designed by \cite{11} and \cite{12}, respectively. By utilizing CS theory, a codebook is designed in \cite{10} for compressed CSI. Similarly, in \cite{13}, an angle-of-departure (AoD)-based codebook design is proposed, which compresses CSI more accurately; but overhead increases linearly with the AoDs. The downside of the codebook is it is environment-dependent. 

Motivated by ML and its wide applications to solve complex optimization problems, its use has widely been adopted in various aspects of wireless communications \cite{Karam_AI, Jakob, karam_UAVBlockChain, rose, Karam_RNN_Twin}. 
Similarly, deep learning (DL), a class of ML algorithms having multiple hidden layers, is utilized for the design of feedback mechanism \cite{DeepL_FB, DeepL_FB2, DeepL_FB3, DeepL_FB4, DeepL_FB5}. In \cite{DeepL_FB}, the CSI matrix is considered as an image and then a DL-based encoder and decoder is designed, named CsiNet, to reduce feedback overhead. Precisely, encoder is used at the UE side to convert the CSI matrix into a codeword, which is then decoded at the gNB by utilizing a decoder. In \cite{DeepL_FB4}, a neural network architecture is proposed, which consists of convolutional neural network-based encoder and decoder, and it process real and imaginary parts of CSI separately. A more practical approach, i.e., compression of noisy (estimation noise) CSI is studied in \cite{DeepL_FB5}, where noisy features are first removed and then compression is performed. All the DL-based approaches treat CSI matrix as an image and then encoders and decoders are designed, where the computation cost of network entities is ignored. Further, to train the deep networks, a massive amount of synthetic data-set is utilized.   

Different from the existing DL-based strategies, i.e., designing encoders and decoders, and accordingly transforming CSI matrix into an image, we introduce the concept\footnote{The proposed idea is tested using MATLAB-based simulations and some initial results are published in IEEE international conference on communications (ICC)'21 \cite{karam_novel, karam_RNN}.} of using channel prediction for CSI feedback (from CSI compression to recovery). Particularly, we exploit twin channel predictors at the gNB and UE for CSI feedback. We call these predictors twin because they use same training/test data-set, same initialization steps, i.e., weights, optimization function, etc. The purpose of a channel predictor is to forecast CSI realizations. For channel prediction, two statistical algorithms, i.e., autoregressive (AR) and parametric models, have been addressed in \cite{KF_Prediction} and \cite{Parametric_Prediction}, respectively. Both models are based on the statistical modeling of a wireless channel. Broadly, a parametric model is a process of estimating Doppler shift, the number of scattering resources, angles of arrival and departure, and amplitude. The assumption is that a wireless channel is a superposition of a finite number of complex sinusoids. Finally, CSI is predicted based on estimated parameters. However, estimated parameters can expire quickly in a highly dynamic environment; therefore, iterative re-estimation of parameters is required, increasing computation cost  \cite{Parametric_Prediction}. In contrast, AR model approximates the wireless channel as an AR process \cite{AR_model}, where the future CSI is extrapolated using a weighted linear combination of current and past CSI. The downside of AR is its vulnerability to impairments, e.g., additive noise, making the AR model infeasible. Therefore, seeing the promising benefits of ML in various domains, e.g., natural language processing, 
we utilize ML for CSI prediction \cite{Karam_CP}.  


 \subsection{Our Contributions}
 This paper provides a pragmatic solution for CSI compression and its recovery by exploiting ML. The major contributions of this paper are summarized as follows.
 
By considering the realistic scenario, we propose the idea of using twin channel predictors, called as hybrid approach, for CSI feedback. The proposed idea is completely different from the traditional concept, i.e, using encoders and decoders, for CSI feedback. More specifically, the proposed approach is hybrid, i.e, it is a combination of conventional CSI feedback and ML-based twin predictors. The hybrid approach can help to eliminate the feedback if the prediction is aligned with the estimation; thus, avoiding feedback overhead. Alternatively, it reaps benefits of reducing feedback overhead and CSI compression errors.
  
Given the realistic CSI information, the training of twin ML predictors is based on the true labels, which exist in the realistic environments, rather than theoretical assumptions as followed for the design of CSI encoders and decoders. Further, the utilization of channel predictors can help to support the possible standardization in third generation partnership project (3GPP). Therefore, we also cover the possible standardization points of the proposed hybrid approach. 

To corroborate the validity of the proposed hybrid approach, we use the experimental data, which is recorded at the Nokia Bell-Labs campus in Stuttgart, Germany. To the best of our knowledge, this is the first work that considers the use of measurement data in the domain of CSI feedback, both for conventional CSI acquisition and validation of the proposed hybrid approach.    

 \subsection{Paper Organization and Notations}

The rest of the paper is organized as follows. The system model is presented in Section\,\ref{system_model}. Conventional CSI feedback mechanism is summarized in Section\,\ref{Conventional Method}. In Section\,\ref{hybrid_approach}, the proposed hybrid approach, the design of channel predictor, and remarks related to hybrid approach are given. The measurement campaign is summarized in Section\,\ref{measurement_campain}. Numerical results and their analysis are presented in Section\,\ref{results}. Finally, in Section\,\ref{conclusion}, conclusions and future outlooks are drawn.

Throughout this paper, the matrices and vectors are represented by boldface upper and lower-case, respectively. Also, scalars are denoted by normal lower-case. The estimated, quantized estimated, predicted, and true channel matrix are represented by $\widehat{\mathbf{H}}$, $\widehat{\mathbf{H}}^{Q}$, $\widetilde{\mathbf{H}}$, and $\mathbf{H}$, respectively, and their vectorized forms are respectively denoted by $\widehat{\mathbf{h}}$, $\widehat{\mathbf{h}}^{Q}$, $\widetilde{\mathbf{h}}$, and $\mathbf{h}$. 
In addition, $Q_c (\cdot)$ and $Q_h (\cdot)$ represent the quantization functions for the conventional and the hybrid approach, respectively. The superscripts $[\cdot]^{T}$ and $(\cdot)^{*}$ denote the transpose and conjugate transpose of a matrix/vector, respectively. Furthermore, $\mathsf{E}\{\cdot\}$ denotes expectation operator and $\left\|\cdot\right\|^2_\text{FRO}$ is the squared Frobenius norm.

\section{System Model}\label{system_model}
\begin{figure}[]
\begin{center}
 \includegraphics[width=8.8cm,height=6cm]{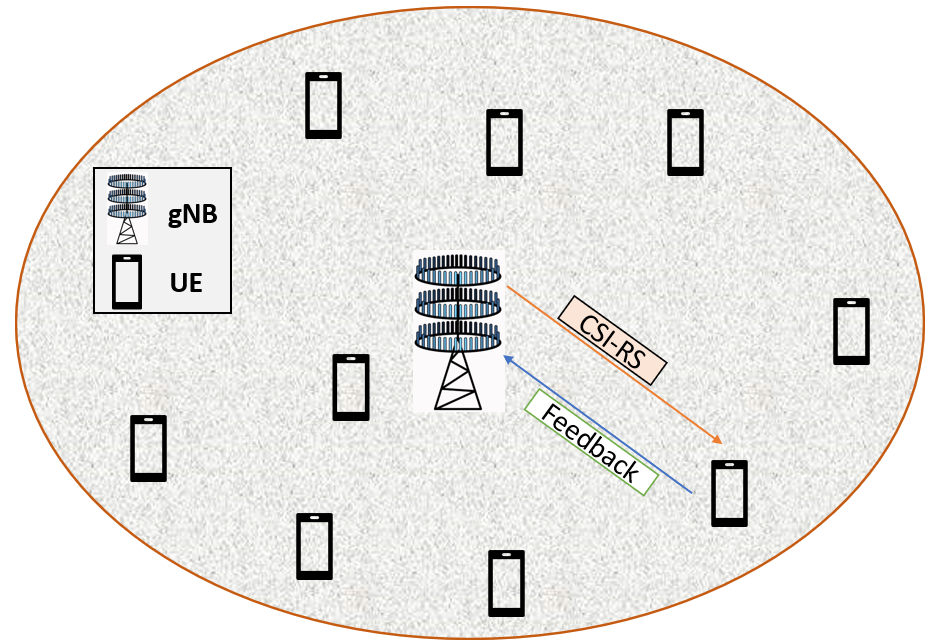}
  \caption{{Single-cell mMIMO-OFDM communication environment, where two network entities are depicted, i.e., a gNB and a UE. To acquire the channel at gNB, gNB is sending CSI-RS; consequently, compressed estimated channel is fed back via dedicated feedback link}.}\label{fig1}      
\end{center}
\end{figure}
Consider the downlink of a mMIMO orthogonal frequency-division multiplexing (OFDM), mMIMO-OFDM, system, as depicted in Fig.\,\ref{fig1}, where a gNB is serving multiple UEs within a cell. For the sake of simplicity, we explain the communication between a single gNB-UE link. The gNB and UE are equipped with $N_t$ and $N_r$ transmit and receive antennas, respectively. In the downlink scenario, the UE estimates the channel through dedicated reference symbols (RSs) and feedback the compressed estimated channel to the gNB. Without loss of generality, the received signal\footnote{For the reader's understanding, here we are referring to one OFDM symbol and subcarrier.} at the UE can be represented as
\begin{equation}\label{MIMO_model}
    \mathbf{y}(t)=\mathbf{H}(t)\cdot\mathbf{p}(t)+ \mathbf{n}(t)\:
\end{equation}
where $\mathbf{y}(t)=\big[y_{1}(t), y_{2}(t), \cdots, y_{N_{r}}(t)\big]^{T}$ having dimension $N_{r}\times 1$, and $t$ is the time index. And,
\begin{gather}
{\mathbf{H}}(t) =
  \begin{bmatrix}
    {h}_{11}(t)&\cdots & {h}_{1N_t}(t)\\
    {h}_{21}(t)&\cdots & {h}_{2N_t}(t)\\
    \vdots&\ddots&\vdots\\
    {h}_{{N_r1}}(t)&\cdots &{h}_{N_rN_t}(t)
   \end{bmatrix}\:
\end{gather}
is the channel matrix, which has dimension $N_{r}\times N_{t}$. In the above matrix, $h_{n_{r}n_{t}}\in \mathbb{C}^{1\times1}$ is the channel gain between the $n_{t}^{th}$ transmit and $n_{r}^{th}$ receive antennas, respectively, where $1\leq n_{t}\leq N_{t}$ and $1\leq n_{r}\leq N_{r}$. Further, $\mathbf{p}(t)=\big[p_{1}(t), p_{2}(t), \cdots, p_{N_{t}}(t)\big]^{T}$ is an $N_t\times1$ symbol vector transmitted by the $N_t$ antennas, and $\mathbf{n}(t)=\big[n_{1}(t), n_{2}(t), \cdots, n_{N_{r}}(t)\big]^{T}$ is the additive noise vector, which has dimension $N_r\times1$.


In the following, we summarize CSI acquisition scheme followed in the conventional approach. Then, we will explain the proposed hybrid approach, which is the combination of conventional approach and ML-based channel predictors. 
\section{Conventional Approach}\label{Conventional Method}
Conventionally, CSI at the gNB is acquired through dedicated RSs. The gNB transmits CSI-RSs, and then the UE estimates the channel by using a specified channel estimator, e.g., linear minimum mean-squared-error (LMMSE) estimator \cite{LMMSE}. 
Consequently, the UE estimates the channel, denoted by $\widehat{\mathbf{H}}_{\text{UE}}$, and then feedback a compressed estimated channel to reduce overhead. This compression is based on a standard element-wise quantization function, denoted by $Q_c(\cdot)$, where real and imaginary parts of the channel are separately quantized. The compressed estimated channel at the UE, at time $t$, can be written as 
\begin{equation}
    \widehat{\mathbf{H}}^{Q}_\text{UE}(t)= Q_c\left(\widehat{\mathbf{H}}_\text{{UE}}(t)\right)\:.
    \label{conventional_sequation}
\end{equation}
The compressed channel is fed back to the gNB. For brevity, we rewrite the acquired channel at the gNB and UE as
\begin{equation}
    \widehat{\mathbf{H}}^{Q}_{\text{}}(t)= Q_c\left(\widehat{\mathbf{H}}_\text{UE}(t)\right)\:.
    \label{c_P_final}
\end{equation}

It can be observed from Equation\,\eqref{c_P_final} that the acquired channel at the gNB is degraded by two major factors. The first is caused by estimation error and the second by \textit{compression} or say \textit{quantization}. Throughout the paper, we use these two terms interchangeably. 
To overcome the first error, i.e., error caused by estimation, is another research topic\footnote{To read the problem of channel estimation, interested readers can refer to, e.g., \cite{LMMSE}.}. In this study, we do not consider the problem of channel estimation. Therefore, in this work, we assume that there is no estimation error. Motivated by the errors caused by \textit{compression}, we present a hybrid approach, explained below.

\section{Proposed Hybrid Approach}\label{hybrid_approach}

\begin{figure}[]
\begin{center}
  \includegraphics[width=8.8cm,height=6.5cm]{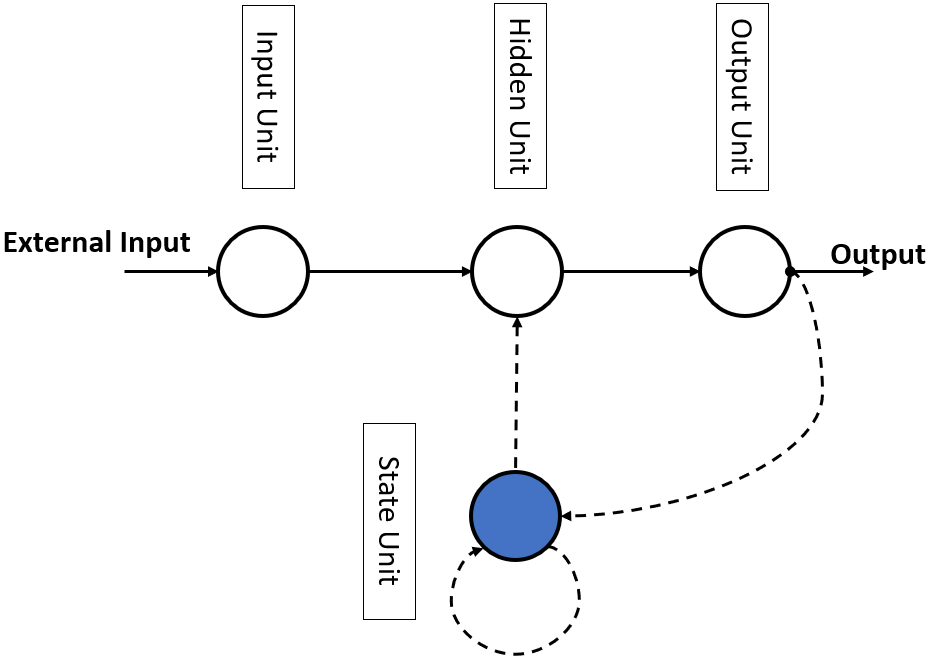}
  \caption{{The topology of a simple \textit{Jordan} RNN \cite{jordan}}.}\label{Jordan}      
\end{center}
\end{figure}
\begin{figure*}[ht]
\begin{center}
  \includegraphics[width=14cm,height=7cm]{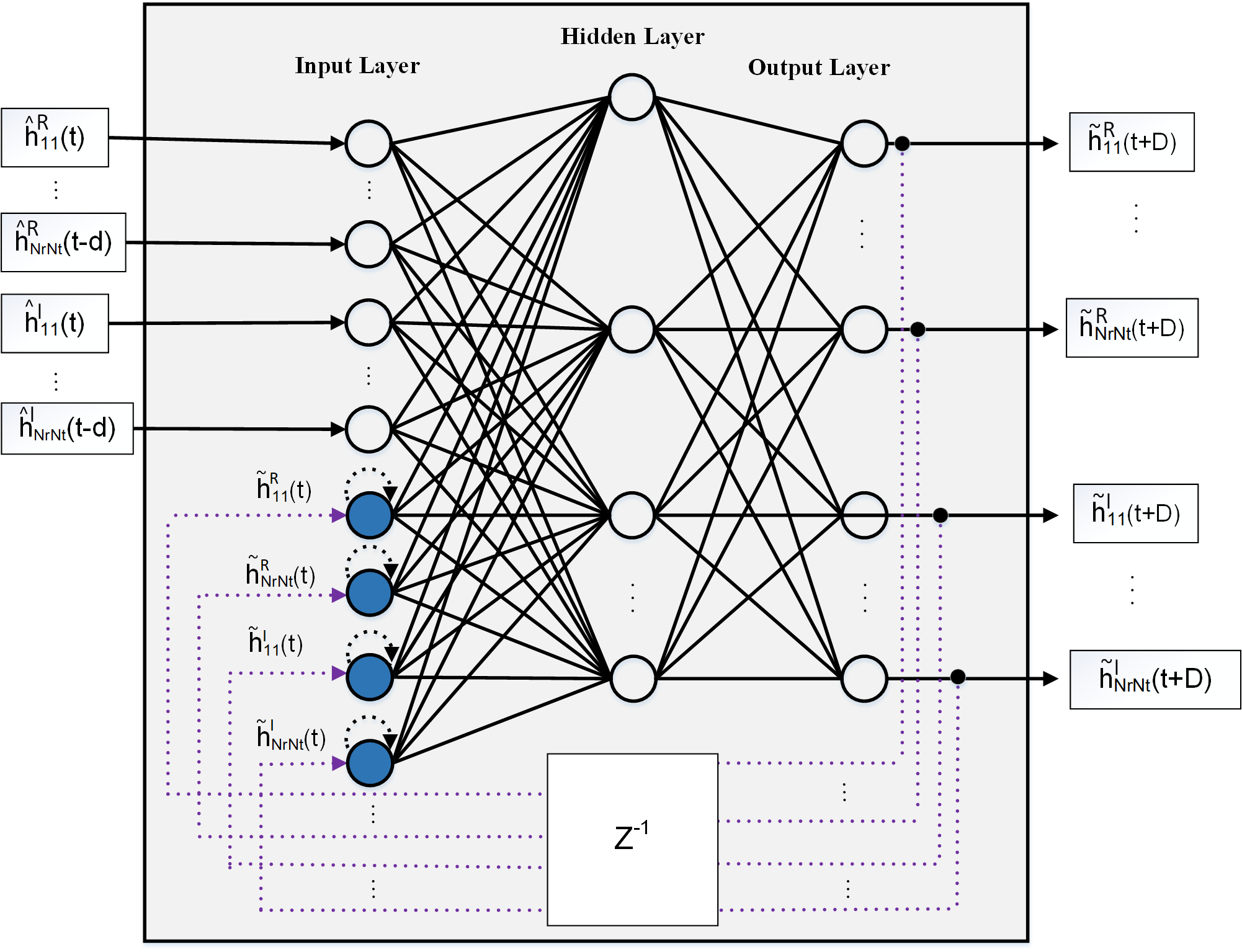}
  \caption{{Pictorial representation of a  fully connected \textit{Jordan} RNN for mMIMO channel prediction. The external inputs of the RNN are the real and imaginary parts of the compressed CSI to be predicted whereas the internal inputs (denoted with blue units) are the recurrent components. The output shows the multi-step predicted real and imaginary parts of compressed CSI for each antenna element.}}\label{RNN}      
\end{center}
\end{figure*}
For the sake of simplicity, we explain the proposed hybrid approach\footnote{The proposed hybrid approach can work in TDD assuming channel reciprocity.} using a single gNB-UE link, remarking that the same communication process would be followed by other UEs present within the network. The hybrid approach considers the use of twin channel predictors at both ends of the communication system, i.e., gNB and UE. The UE will evaluate the feedback with respect to prediction. As a toy example, if there is no difference between the predicted and the estimated channel at the UE, then feedback is not required. 
The hybrid approach is based on channel prediction; therefore, we first summarize ML-based channel predictor below \cite{Karam_CP}. Then, we will explain the hybrid approach.   
\subsection{ML-Based mMIMO Channel
Predictor}\label{channel_predictor}

With the rapid advancements in ML algorithms, their applications have been considered in almost every field. For example, 
recurrent neural network (RNN), a class of artificial neural network (ANN), is a powerful technique for time-series predictions \cite{RNN_paper}. RNNs are different from basic neural networks (e.g., feed-forward neural networks) since their training and prediction are based on current and past information. 
Contrarily, a basic neural network's output is independent of previous input, which may not be the case in many scenarios, e.g., consider correlated fading channel in wireless communications. Seeing the promising benefits of RNN, we also use RNN as an mMIMO channel predictor in this study \cite{Karam_CP}. 

The RNN has several variants 
among which we use a \textit{Jordan network} \cite{jordan}. The basic topology of a \textit{Jordan network} is drawn in Fig.\,\ref{Jordan}. 
The layers are connected in a feed-forward configuration\footnote{It is important to note that \textit{Jordan} used the terms plan unit and state unit for external and internal input, respectively, in his paper \cite{jordan}. However, for the sake of the reader's understanding, we use them as external and internal input.}. The hidden and output layers are also connected with a state unit (we call them internal input); this recurrent connection of the internal input provides a short-term memory to the hidden unit. Thereby, the hidden unit is influenced by the external input and their past state obtaining via internal input. The addition of an internal input helps the RNN find a time-series relationship between the various inputs; thus, making it different from a basic neural network.
\begin{figure*}
\begin{center}

  \includegraphics[width=18cm,height=3cm]{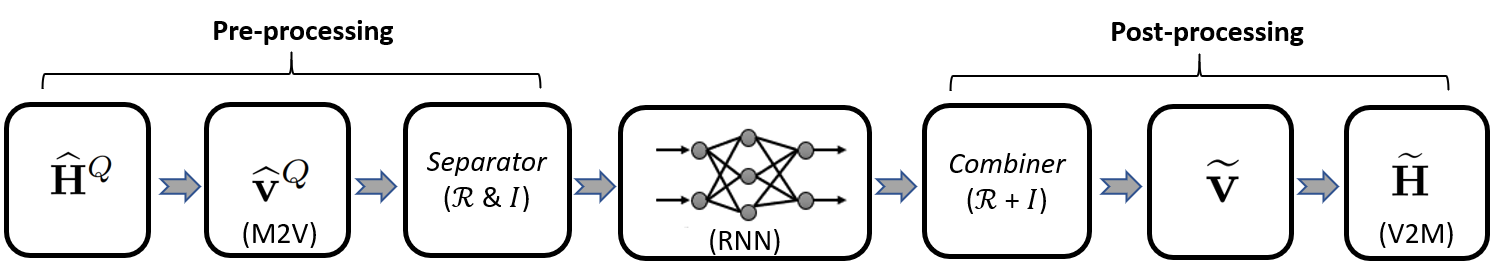}
  \caption{\textcolor{black}{Pre and post-processing of CSI matrix for training and prediction using the RNN. The real and imaginary parts of CSI are represented by $\mathcal{R}$ and ${I}$, respectively.}}\label{processing}

\end{center}
\end{figure*}

By following the same structure as of \textit{Jordan network}, we reconstructed\footnote{A similar configuration is also followed in \cite{RNN_VTC} for single-input single-output configuration.} a mMIMO RNN, as drawn in Fig.\,\ref{RNN}, for the prediction of CSI \cite{Karam_CP}. The objective of the RNN is to forecast multi-step ahead values of CSI, denoted by $\widetilde{\mathbf{H}}(t+D)$, which are as close as possible to true values, i.e., $\widehat{\mathbf{H}}^{Q}(t+D)$. We assume the total (external and internal) numbers of neurons in the input layer as $I$, hidden layer as $J$, and output layer as $K$. In terms of mMIMO channel prediction, the input of the RNN is the compressed channel, written as
\begin{gather}
\widehat{\mathbf{H}}^{Q}(t) =
  \begin{bmatrix}
    \widehat{h}^{Q}_{11}(t)&\cdots & \widehat{h}^{Q}_{1N_t}(t)\\
    \widehat{h}^{Q}_{21}(t)&\cdots & \widehat{h}^{Q}_{2N_t}(t)\\
    \vdots&\ddots&\vdots\\
    \widehat{h}^{Q}_{N_r1}(t)&\cdots &\widehat{h}^{Q}_{N_rN_t}(t)
   \end{bmatrix}\:
\end{gather}
where $\widehat{h}^{Q}_{n_rn_t}(t)\in \mathbb{C}^{1\times1}$ is the compressed estimated channel between $n_t^{th}$ and $n_r^{th}$ transmit and receive antennas, respectively. Along with the compressed channel, $\widehat{\mathbf{H}}^{Q}(t)$, its $d\text{-step}$ delayed versions, denoted by $[\widehat{\mathbf{H}}^{Q}(t-1), \widehat{\mathbf{H}}^{Q}(t-2), \cdots, \widehat{\mathbf{H}}^{Q}(t-d)]$, are also fed as an external input to the RNN. RNN's total external input is given as
\begin{equation}
    \widehat{\mathbf{H}}^{Q}=\big[\widehat{\mathbf{H}}^{Q}(t),\widehat{\mathbf{H}}^{Q}(t-1), \widehat{\mathbf{H}}^{Q}(t-2), \cdots, \widehat{\mathbf{H}}^{Q}(t-d)\big].
\end{equation}
The classic RNN do not take matrices as input,
hence pre-processing is required. Below we provide some pre-processing steps, which are also depicted in Fig.\,\ref{processing}. First, the combined input $\widehat{\mathbf{H}}^{Q}$ is converted from matrix-to-vector (M2V) form as
\begin{equation}
    \widehat{\mathbf{v}}^Q= \textrm{M2V}(\widehat{\mathbf{H}}^{Q})= \big[\widehat{h}_{11}^Q,\quad \widehat{h}_{12}^Q, \quad\cdots,\quad \widehat{h}_{N_rN_t}^Q\big]\:
\end{equation}
which represents a vector having complex entries. Currently, the ML algorithms are not well implemented for complex values \cite{complex_NNs}; therefore, in our study, we use real-valued RNN. For this purpose, the above complex vector can be converted into separate real and imaginary parts through a dedicated function, we call it \textit{separator}; the outcome can be represented into a vector form as

\begin{equation}
    {\widehat{\mathbf{v}}}= \big[{\widehat{h}}_{11}^R(t),\quad \cdots,\quad  \widehat{h}_{N_rN_t}^R(t-d),\: \widehat{h}_{11}^I(t), \quad\cdots,\quad \widehat{h}_{N_rN_t}^I(t-d)\big]\:
\end{equation}
where $\widehat{h}_{n_rn_t}^R$ and $\widehat{h}_{n_rn_t}^I$ denote\footnote{For notational convenience, we removed the superscript $Q$, which depicts the compressed channel.} the real and imaginary parts of a complex channel $\widehat{h}^{Q}_{n_rn_t}\in \mathbb{C}^{1\times1}$. These inputs can be seen in Fig.\,\ref{RNN}, where real and imaginary parts are fed as an external input to the RNN. 

Besides, the recurrent\footnote{To get the deeper understanding of the recurrent block, the interested readers can refer to \cite{jordan}.} (feedback or internal inputs) components for time $t$, as depicted with blue units (or neurons) in Fig.\,\ref{RNN}, are expressed in the vector form as
\begin{equation}
    \widetilde{\mathbf{v}}(t)= \big[\widetilde{h}_{11}^R(t),\quad \cdots,\quad  \widetilde{h}_{N_rN_t}^R(t),\: \widetilde{h}_{11}^I(t), \quad\cdots,\quad \widetilde{h}_{N_rN_t}^I(t)\big]\:
\end{equation}
where $\widetilde{h}_{n_rn_t}^{R}(t)$ and $\widetilde{h}_{n_rn_t}^{I}(t)$ show the real and imaginary parts of the channel, $\widetilde{h}_{n_rn_t}(t)\in \mathbb{C}^{1\times1}$, predicted at time $t$, which acts as a recurrent component for the next iteration. The above vector is fed into RNN internally, which can be seen in the form of feedback in Fig.\,\ref{RNN}. The total input, i.e., external and internal, of the RNN can be represented in a vector form as
\begin{equation}
    {\mathbf{q}}(t)= \big[\widehat{\mathbf{v}}, \widetilde{\mathbf{v}}(t)\big]\:.
\end{equation}
In  a nutshell, by feeding the total input into the RNN, the $D$-step ahead prediction of CSI, can be written as

\begin{equation} 
\begin{split}
    \widetilde{\mathbf{h}}^{R,I}(t+D) &= \big[\widetilde{h}_{11}^R(t+D),\quad \cdots\quad  ,\widetilde{h}_{N_rN_t}^R(t+D),\\
    &\quad\:\:\,\widetilde{h}_{11}^I(t+D), \quad\cdots\quad ,\widetilde{h}_{N_rN_t}^I(t+D)\big]\:.
\end{split}
\end{equation}
Next, by applying the post-processing steps, shown in Fig.\,\ref{processing}, first we combine the real and imaginary parts using a function, called \textit{combiner}. Hence, the complex predicted channel vector is given as
\begin{equation} 
    \widetilde{\mathbf{h}}(t+D) = \big[\widetilde{h}_{11}(t+D),\quad\cdots\quad  ,\widetilde{h}_{N_rN_t}(t+D)\big]\:.
\end{equation}
Lastly, by performing vector-to-matrix (V2M) conversion, the above vector can be written into the matrix form as
\begin{gather}
\widetilde{\mathbf{H}}(t+D) =
  \begin{bmatrix}
    \widetilde{h}_{11}(t+D)&\cdots & \widetilde{h}_{1N_t}(t+D)\\
    \widetilde{h}_{21}(t+D)&\cdots & \widetilde{h}_{2N_t}(t+D)\\
    \vdots&\ddots&\vdots\\
    \widetilde{h}_{N_r1}(t+D)&\cdots &\widetilde{h}_{N_rN_t}(t+D)
   \end{bmatrix}\:
\end{gather}
which represents the mMIMO predicted channel at time step $t+D$. Besides, $\widetilde{h}_{n_rn_t}(t+D)\in \mathbb{C}^{1\times1}$ denotes the predicted channel for the $n_t^{th}$ transmit and $n_r^{th}$ receive antennas.

The prediction of the RNN is based on the weights (generally denoted with $w$) of input and hidden layers and the applied activation function to the linear equation. In summary,
the weight, $w$, is randomly chosen for the connection between the output of a neuron in the predecessor layer and the input of a neuron in the successor layer. Let us denote the weight of $i^{th}$ input and $j^{th}$ hidden neuron as $w_{ji}$, and $w_{kj}$ depicts the weight for the connection between $k^{th}$ output neuron and $j^{th}$ hidden neuron, where $1 \leq i\leq I$, $1 \leq j\leq J$, and $1\leq k\leq K$. The output of a neuron is calculated by applying the activation function to the linear equation (combination of input and weight value); such activation function introduces non-linearity. Generally, there are four common activation functions, which are used in the ML domain, namely sigmoid, hyperbolic tangent, linear, rectified linear, leaky rectified linear, and threshold. Through experiments, we found that the hyperbolic tangent function gives the best result. 
The chosen activation function is expressed as
\begin{equation}
    tanh(\beta)=\frac{e^{\beta}-e^{-\beta}}{e^{\beta}+e^{-\beta}}\label{act_func1}\:.
\end{equation}
where $\beta$ is the value on which activation is applied and it can be written as
\begin{equation}
    \beta=\mathbf{w}_j\cdot\mathbf{q}(t)\:
\end{equation}
where $\mathbf{w}_j= [w_{j1}, \cdots,w_{jJ}]$ represents the weights of hidden layer. The output of the $j^{th}$ hidden layer neuron, at time instant $t$, can be calculated as
\begin{equation}
    o_{j}(t)=tanh\left(\mathbf{w}_j\cdot\mathbf{q}(t)\right)\:.\label{act}
\end{equation}
Consequently, the above activation function is send as an input to the next layer, which in our case, is output layer (see Fig.\,\ref{RNN}). In summary, the $D\text{-step}$ ahead predicted output at the $k^{th}$ output neuron can be written as 
\begin{equation}
    O^k(t+D)=\sum_{j=1}^{J}w_{kj}\cdot o_j(t)\:.
    \label{output}
\end{equation}
In mMIMO RNN model, the above output represents the real or imaginary part of the $n_t^{th}$ and $n_r^{th}$ transmit and receive antennas, respectively.

The operation of an RNN is divided into two phases: training and prediction. In the first phase, after selecting the hyperparameters (e.g., learning rate, optimization algorithm), the training phase can be started. In this phase, channel values are given at input of RNN, along with corresponding output labels. Consequently, the RNN processes each sample (or a batch of samples depending on the batch size), and compares the predicted value with the true label. The weights are updated through backpropagation, which is repeated until a certain convergence condition, e.g., mean-squared-error (MSE) is minimized. The trained RNN can make multi-step ahead prediction. The readers, who want to get further understanding of the working of RNN and its model, are referred to read, e.g., \cite{Karam_CP, jordan}.
\subsection{CSI Feedback with the Aid of RNN}
In the following, we explain the proposed hybrid approach with the aid of an RNN-based channel predictor. In summary, the hybrid approach consists of the following phases: assessment, initialization, prediction, estimation, compression, feedback and recovery of compressed CSI. Below, we describe these phases in more detail.
\subsubsection{Assessment Phase}
\begin{figure}[]
\begin{center}
  \includegraphics[width=8.8cm,height=5.5cm]{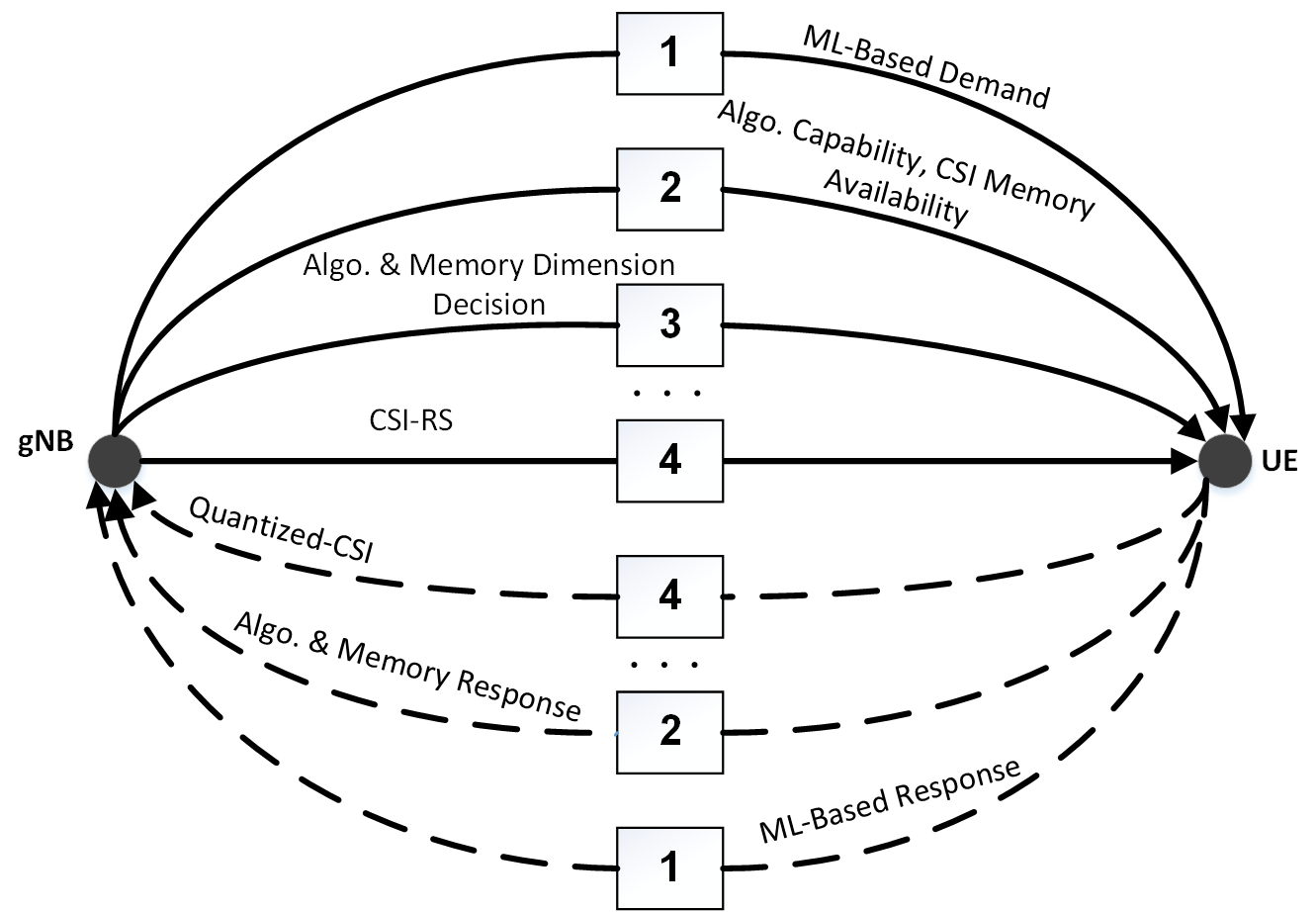}
  \caption{Dedicated message exchanges, we call this as assessment phase, between the gNB and UE before adopting the hybrid approach.}\label{assesment_phase}      
\end{center}
\end{figure}
To adopt the hybrid approach, the two network entities exchange a few messages: 1) gNB and UE perform a handshake to assess available capabilities, for instance, support of ML algorithm 2) both agree to use ML-based algorithm, e.g., in our work, RNN 3) both agree to adopt the conventional approach to aggregate past CSI realizations to train the RNN 4) the length of the data-set (to be acquired in the initialization phase) is decided for the training of RNN. The assessment phase is pictorially represented in Fig.\,\ref{assesment_phase}. Once such messages are exchanged, the gNB and UE perform the initialization phase, which is summarized below.
\subsubsection{Initialization Phase}
During the initialization phase, the gNB and UE use the conventional approach, detailed in Section\,\ref{Conventional Method}. In summary, the gNB sends CSI-RSs to acquire the channel at time $t$; in consequence, the UE feedback compressed estimated channel, denoted by $\widehat{\mathbf{H}}^{Q}(t)$. By using conventional approach, a set of past CSI realizations is established at both ends, where the length of data-set is agreed in the assessment phase (see point $4$). Let us denote the set of past CSI realizations as $\mathcal{S} \delequal \{\widehat{\mathbf{H}}^{Q}(t), \widehat{\mathbf{H}}^{Q}(t-1), \widehat{\mathbf{H}}^{Q}(t-2), \cdots, \widehat{\mathbf{H}}^{Q}(t-d), \cdots,  \widehat{\mathbf{H}}^{Q}(t-S)\}$, where $S$ is the total length of initialization phase.
\subsubsection{Prediction Phase}
By using the same data-set, available at the gNB and UE, both the network entities train the RNN, which is already explained in Section\,\ref{channel_predictor}. 
By having the trained RNNs at both sides, let us say at time instant $t$, the gNB and UE predict the channel, denoted by $\widetilde{\mathbf{H}} (t)$. The prediction is assumed to be the same on both sides. For the sake of notational convenience, we rewrite the predicted channel at gNB and UE as $\widetilde{\mathbf{H}}_{\text{gNB}} (t)$ and $\widetilde{\mathbf{H}}_\text{UE} (t)$, respectively.
\subsubsection{Estimation Phase}
In this phase, the UE estimates the channel using a dedicated channel estimator, e.g., LMMSE. Suppose that the UE estimates the channel at time $t$, which is denoted as $\widehat{\mathbf{H}}_\text{UE}(t)$. In our study, we assume that there is no error in the estimation; therefore, the estimated channel will be the same irrespective of the algorithm being used for channel estimation. The reason for considering zero-estimation error is to observe only the errors caused by compression. 
\subsubsection{Compression Phase}
Briefly, in the conventional approach, compression at the UE will be done in the same manner as explained in Section\,\ref{Conventional Method}, i.e., Equation\,\eqref{conventional_sequation}. Whereas in the hybrid approach, compression at the UE is done in the following manner. 

The update function, by exploiting predicted and estimated channel, at the UE is computed as
\begin{equation}\label{update}
    \mathbf{U}(t)=f\left(\widetilde{\mathbf{H}}_\text{UE} (t), \widehat{\mathbf{H}}_\text{UE} (t)\right)
\end{equation}
where $f(\cdot)$ is an update function, which measures the distance between the predicted and estimated channel at UE. A simple implementation of such a function could be a difference. Hence, Equation \eqref{update} can be rewritten as
\begin{equation}
    \mathbf{D}(t)=\widetilde{\mathbf{H}}_\text{UE} (t) - \widehat{\mathbf{H}}_\text{UE} (t)\:.
    \label{updateq}
\end{equation}
In succession, the updated function, $\mathbf{D}(t)$, is compressed at the UE using a standard element-wise quantization scheme. The compressed channel at the UE can be written as
\begin{equation} 
    \widehat{\mathbf{H}}_h^{Q}(t) = Q_h\left(\mathbf{D}(t)\right)\:
     \label{updateq2}
\end{equation}
where $Q_h(\cdot)$ is the quantization function. By substituting Equation\,\eqref{updateq} into Equation\,\eqref{updateq2}, we obtain the compressed channel as
\begin{equation} 
    \widehat{\mathbf{H}}^{Q}_h(t) = Q_h\left(\widetilde{\mathbf{H}}_\text{UE} (t) - \widehat{\mathbf{H}}_\text{UE} (t)\right)\:.
     \label{updateq2_i}
\end{equation}

\subsubsection{Feedback and Recovery Phase}
The feedback and correspondingly the recovery of the compressed channel is categorized into two scenarios. In the following, we explain each scenario in detail.
\begin{itemize}
    \item In the first scenario, we consider the case of perfect channel prediction, i.e., when the estimated and predicted channel at the UE are same. In other words, $\widetilde{\mathbf{H}}_\text{UE} (t) = \widehat{\mathbf{H}}_\text{UE} (t)$. In this case, the UE does not need to feedback anything, and hence, feedbcak-related overhead is eliminated. Notably, the acquired channel at the gNB will be the same, which it had predicted, i.e., $\widetilde{\mathbf{H}}_\text{gNB} (t)$. Mathematically, the acquired channel at the gNB is given as
\begin{equation} 
    \widehat{\mathbf{H}}_\text{gNB}(t) = \widetilde{\mathbf{H}}_\text{gNB}(t)\:.
\end{equation}
\item In the second scenario, we consider the case when the UE has to feedback something. In short, when $\widetilde{\mathbf{H}}_\text{UE} (t) \neq \widehat{\mathbf{H}}_\text{UE} (t)$. Therefore, the UE feedback the compressed update, i.e., Equation\,\eqref{updateq2_i}. Correspondingly, the acquired channel (or the recovered channel) at the gNB can be calculated as
\begin{equation} 
    \widehat{\mathbf{H}}_\text{gNB}(t) = \widetilde{\mathbf{H}}_\text{gNB}(t)-Q_h\left(\widetilde{\mathbf{H}}_\text{UE} (t) - \widehat{\mathbf{H}}_\text{UE} (t)\right)\:
    \label{final}
\end{equation}
which is simply the difference between the predicted channel at the gNB and feedback from the UE.
\end{itemize}
Based on the hybrid approach, in the following subsection, we provide a few remarks.
\subsection{Remarks}
There are threefold advantages of using the proposed hybrid approach over the conventional approach. Firstly, the errors caused by compression can be smaller in Equation\,\eqref{updateq2_i} than in Equation\,\eqref{conventional_sequation}. Thanks to the fact of having low amplitude entries in Equation\,\eqref{updateq2_i}, which is due to prediction. Therefore, a more accurate version of the true channel can be acquired at the gNB. The second benefit can be attained in terms of quantization bits required to send feedback from the UE. The better the prediction at the UE, the fewer bits would be needed to send the feedback. Thus, Equation\,\eqref{updateq2_i} can significantly reduce feedback bits than Equation\,\eqref{conventional_sequation}. Lastly, if the prediction at the UE is perfect, then feedback is not necessary; hence, feedback-related overhead is completely eliminated. This benefit can be attained in the environment where the UEs are static (e.g., in a stadium, or moving at a predictable speed (in trains etc.)). This is because when the UEs are static, variations in the channel can be very small; hence, prediction can be more accurate.

If the quantization functions, $Q_h(\cdot)$ and $Q_c(\cdot)$, are using an infinite amount of bits to send the feedback, then there is no advantage of using the hybrid approach. In short, quantization followed in the conventional approach and in the hybrid approach will be same, i.e., $Q_c(x)=Q_h(x)=x$, where $x$ is the input data for quantization. Or, we can say that the UE is not making compression errors while sending feedback. Hence, the real benefits can be acquired when the channel is highly compressed, which happens in reality, i.e., type-I and type-II CSI feedback used in 3GPP are highly compressed \cite{Type-I:3GPP}.

The third important remark is related to making a prediction at time $t+1$, i.e., the use of input data at time $t+1$ for the prediction of the next CSI value. Till time $t$, we assumed that $\widehat{\mathbf{H}}^{Q}(t)$ is available to make prediction but at time $t+1$, due to the use of hybrid approach, $\widehat{\mathbf{H}}^{Q}(t)$ is not available at the gNB. Therefore, the two available options at both ends are to either use the update function, i.e., Equation\,\eqref{updateq2_i}, or use Equation\,\eqref{final}. It is verified through experiments\footnote{The results of such verification(s) are not addressed as they are unnecessary.} that the Equation\,\eqref{final} is the best option, which is obvious as this value will be more close to the input data used in training. Also, this concept verifies that the ML algorithms' training and test data-set must come from the same distribution. 

Lastly, as the channel prediction is employed at both ends of the communication system; hence, error may add at both ends if the prediction is bad
. This error can grow with the passage of time as the prediction for $t+T$, where $T$ is the desired time instant for which prediction is required, is dependent on Equation\,\eqref{final}. To overcome this problem, we proposed the hybrid approach, i.e., the conventional approach and the proposed approach will be working together. Let us assume that the prediction at time $t$ is perfect, and in time interval $[t,T]$ error is induced due to bad prediction, and this error reached a threshold $\Lambda$. Suppose that this threshold is the error observed in the conventional approach. Therefore, when the observed error in the hybrid approach reaches the threshold $\Lambda$, then switch to the conventional approach to gather the data for CSI prediction. Once the prediction is good, then switch back to the hybrid approach and so on. 

The pseudo-code of the proposed hybrid approach is given in Algorithm\,\ref{algo1}. Briefly, in the first step of Algorithm\,\ref{algo1}, the data-set is acquired at the gNB and UE using the conventional approach. Then the estimated channel at the gNB is acquired using Equation\,\eqref{final}, which is summarized in step-2 to step-4. Finally, the hybrid approach is followed till the error is less than the threshold, $\Lambda=\mathbf{H}(t)-\widehat{\mathbf{H}}^Q(t)$, where $\mathbf{H}(t)=\widehat{\mathbf{H}}(t)$ as we assume zero-estimation error. Otherwise, step-1 to step-4 are repeated and so on.  

The proposed hybrid approach needs to be supported by the standard as the invention requires the establishment of a protocol between the gNB and UE, which is currently not part of the 3GPP specification. To this end, different prediction algorithms can be tabled and standardized. Furthermore, the possible standardization points include algorithm(s), memory, message exchanges, etc.
\begin{algorithm}[t!]
\label{algo1}
\DontPrintSemicolon
  \KwInput{$\widehat{\mathbf{H}}^Q(t)$}
  \KwOutput{{$\widehat{\mathbf{H}}_\text{gNB}(t) $}}
  {\tcp{\textcolor{black}{Step-1: Aggregate data using Equation\,\eqref{c_P_final}, create a set $\mathcal{S}$, and reverse its order w.r.t. time, i.e., most fresh entry will go at the end. 
  }}}

  \For{$s=1$ \textbf{to} ${S-1}$}
  {
 \textcolor{black}{Train the RNN using the process explained in Section\,\ref{channel_predictor}.\;}
  }
    {\tcp{\textcolor{black}{Step-2: Predict the channel at time $t$: $\widetilde{\mathbf{H}}_\text{gNB}(t)$ and $\widetilde{\mathbf{H}}_\text{UE}(t)$. }} } 
        {\tcp{\textcolor{black}{Step-3: Estimate the channel at UE at time $t$: $\widehat{\mathbf{H}}_\text{UE}(t)$. }} } 
        {\tcp{\textcolor{black}{Step-4: Obtain $\widehat{\mathbf{H}}_\text{gNB}(t)$ using Equation\,\eqref{final}.}} } 
           
             \While{$( \mathbf{H}(t)-\widehat{\mathbf{H}}_\text{gNB}(t)<\mathbf{H}(t)-\widehat{\mathbf{H}}^Q(t))$}
    {
        
        Use hybrid approach, i.e., Equation\,\eqref{final}.}
             \If{$\mathbf{H}(t)-\widehat{\mathbf{H}}_\text{gNB}(t)\geq\mathbf{H}(t)-\widehat{\mathbf{H}}^Q(t)$}
    {
        
Repeat step-1 to step-4.\;
        }
\caption{\textcolor{black}{Proposed Hybrid Approach}}
\end{algorithm}

\begin{figure*}[]
\begin{center}
  \includegraphics[width=18cm]{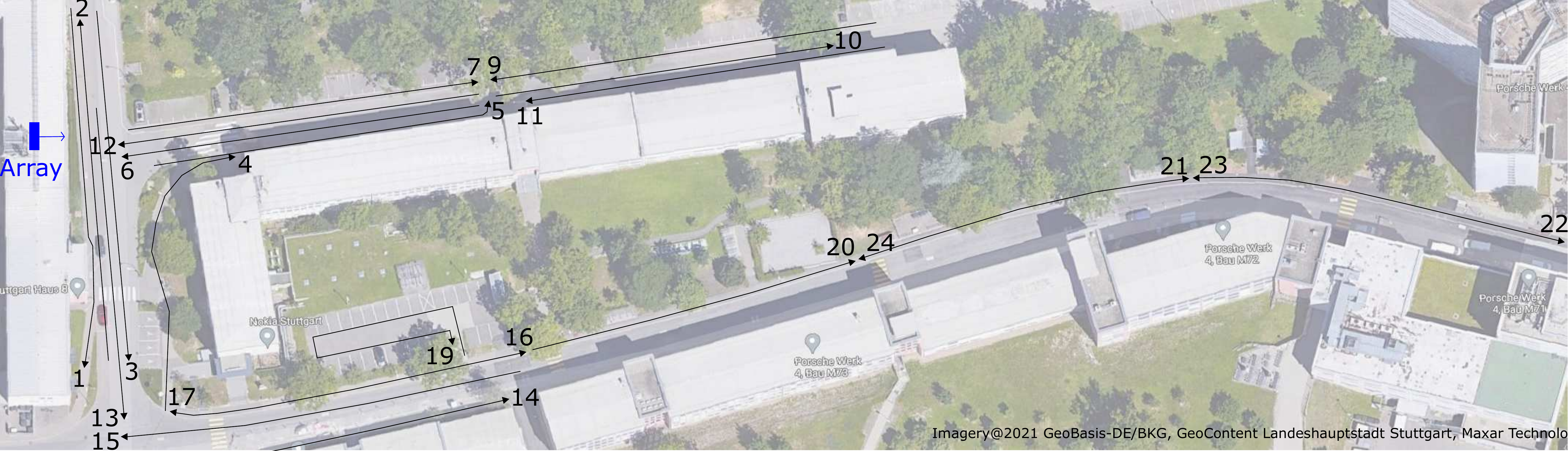}
  \caption{{Image of the Nokia campus in Stuttgart, Germany. The location (resp. boresight direction) of the transmit array is marked on the left by a blue bar (resp. blue arrow). The measurement tracks, along which the vehicle (mimicking a UE) was moved, are depicted by black lines. The corresponding arrowheads and numbers indicate the direction of movement and the measurement track number. Source: \cite{Karam_CP}.}}\label{campaign}      
\end{center}
\end{figure*}

\begin{figure*}
     \centering
     \subfigure[\textcolor{black}{{Track-$1$} of measurement campaign at Nokia Bell-Labs campus in Stuttgart, Germany.}]{\includegraphics[width=8cm,height=5.5cm]{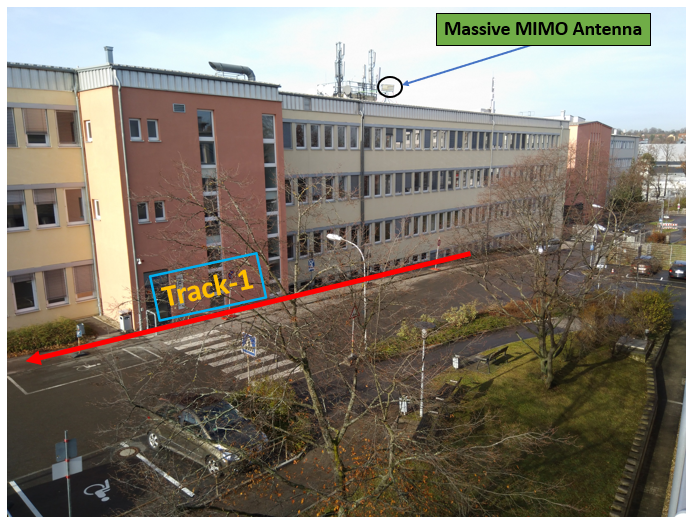}}{\label{route_1}}
     \subfigure[\textcolor{black}{A mMIMO antenna of $64$ antenna elements ($4$ lines and $16$ columns of antenna elements) on the roof of a building.}]{\includegraphics[width=6cm,height=5.5cm]{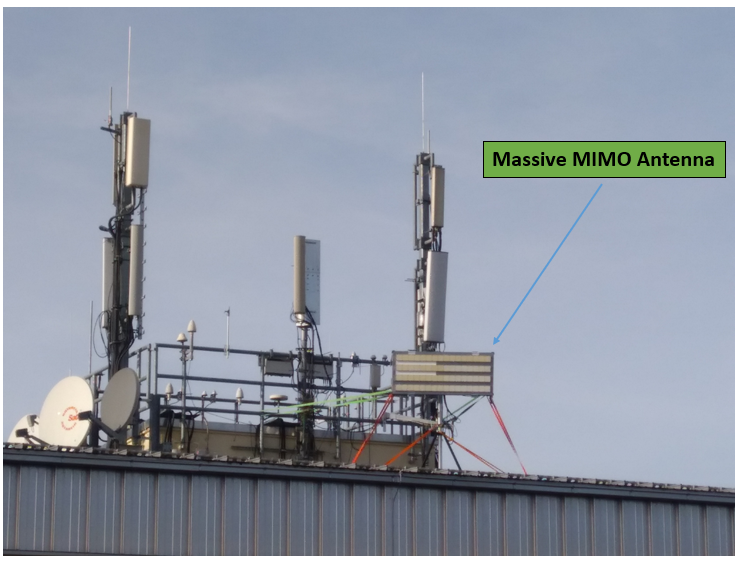}}{\label{antenna_1}}
     \subfigure[\textcolor{black}{A vehicle (mimicking a UE) is moving on the sketched tracks.}]{\includegraphics[width=3cm,height=5.5cm]{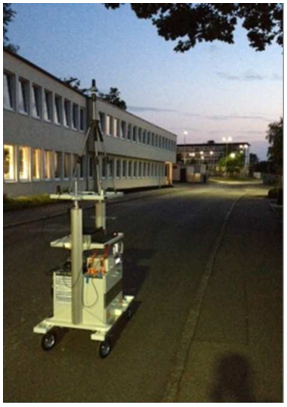}}{\label{UE_1}}
        \caption{\textcolor{black}{Pictorial view of network entities and one of the tracks followed during the measurement campaign.}}
        \label{pictorial_rep}
\end{figure*}

\section{Measurement Campaign}\label{measurement_campain}
A mMIMO channel measurement campaign (see Figs.\,\ref{campaign}\,,\ref{pictorial_rep}) was conducted on the Nokia campus in Stuttgart, Germany \cite{Karam_CP}. In that area, the buildings of approx. $15$ meter height are primarily arranged along streets, acting as reflectors and blockers for the radio waves from the transmit antenna array, which was placed on the roof-top of one of these buildings. The environment resembles well a non-line-of-sight urban-micro-like propagation scenario.
The geometry of the $64$-element transmit array had been adapted to the propagation scenario; that is, $4$ rows with $16$ (single-pol.) patch antennas each were used, with a horizontal antenna spacing of $\lambda/2$, and a vertical separation of $\lambda$.

The array antennas transmitted $64$ time-frequency orthogonal pilot signals at $2.18$\,GHz carrier frequency, using OFDM waveforms that followed the $10$\,MHz LTE numerology (i.e., $600$ subcarriers with $15$\,kHz spacing). The pilot signals had been arranged such that the sounding on $50$ separate subbands (each consisting of $12$ consecutive subcarriers) required $0.5$\,ms. Within that pilot burst period, the propagation channel was assumed to be time invariant. The pilot bursts were sent continuously with a periodicity of $0.5$\,ms.

The receiver cart (mimicking a UE) consisted of a single monopole antenna mounted at $1.5$ meter height, a Rohde \& Schwarz TSMW receiver and a Rohde \& Schwarz IQR hard disc recorder which continuously captured the received baseband signal. Both, the transmit array and the receiver were frequency synchronized via global positioning system (GPS). During the measurements, the receiver cart was moved along several routes at walking speed ($3$ to $5$\,kmh$^{-1}$), which corresponds a spatial channel sampling distance of less than $0.1$\,mm. 
A post-processing step extracted for each pilot burst and subband the $64$-dimensional channel vector. The estimated normalized mean-squared-error (NMSE) in the channel estimates is in the range of $-20$ to $-30$\,dB.

\section{\textcolor{black}{Numerical Results}}\label{results}
In this section, we verify the effectiveness of the proposed approach for CSI feedback. To this end, firstly, measurement scenario is presented. Then we present distribution of the data-set for the training and prediction of the RNN. We demonstrate that the proposed approach outperforms the existing feedback method. The performance is validated using various evaluation parameters including NMSE, precoding gain, spectral efficiency, cosine similarity, and signal-to-noise ratio (SNR). 
\begin{table*} [t]
\caption{Parameters of the RNN and measurement campaign}
\centering
 \begin{tabular}{c c c}
 \hline
 \textbf{Parameter} & \textbf{Value} & \textbf{Description}  \\ 
 \hline
 $v, \gamma$ & $ [3,5]\,\text{kmph}, 40\,\text{dB}$ & Speed of UE, and SNR\\
 $h_t, h_r, f_c$ & $15\,\text{meter}, 1.5\,\text{meter}, 2.18\,\text{GHz}$ & Building's height, receiving antenna's height, and carrier frequency\\ 
 $D_\text{length}, J$&  $116\times10^3, 20$ & Total length of data on each track, and no. of neurons in hidden layer\\
 RNN architecture& $4$-layer& One input, two hidden, and one output layer\\
Optimization algorithm & \textit{Adaptive moment
estimation} (\textit{Adam})  & Algorithm to update weights of RNN\\
 ${D_\text{train}}, {D_\text{valid}}, {D_\text{test}}$ & $80\%, 10\%,10\%$&Distribution of data-set on \textit{track-$1$}\\
 $B_\text{size}$, $l_r$ & $20, 1\times 10^{-4}$&Batch size, and learning rate for training the RNN\\
ML implementation platform& \textit{Tensorflow} with \textit{Python}&The utilized platform for the implementation of RNN \\
  $tanh(\cdot)$, $N_e$ & $\text{Hyperbolic tangent function}, 100$&Activation function, and number of training episodes\\
 \hline
\end{tabular}
\label{table:I}
\end{table*}

\subsection{\textcolor{black}{Scenario and Training Data}}
We consider a single-user (SU)-mMIMO communication environment, where a UE is moving on different tracks, as indicated in Fig.\,\ref{campaign}. The gNB and UE are supported with an RNN-based mMIMO channel predictor. The UE is moving on the tracks with a speed of $3$ to $5\,\text{kmh}^{-1}$. The gNB is equipped with $64$ transmit antennas, and the UE has $1$ receive antenna. The rest of the simulation parameters are summarized in Table\,\ref{table:I}. 

For the training and prediction using the RNN at the gNB and UE, we extracted the data-set of the first four transmit antennas. Importantly, for the training purpose, we used the data-set of track-$1$ only, comprised of $58$ seconds. \textcolor{black}{The measurements provide $2000$ mMIMO channel samples per second}. The $80\%$ data-set of track-$1$ is used for training, $10\%$ for the validation and remaining for the test. Similarly, $10\%$ data-set of track-$2$ and track-$19$ is taken randomly for the test purpose. Numerically, we had a total data-set of track-$1$, comprised of consecutive $116$k data blocks, i.e., ${\mathbf{H}}(t)|1\leq t\leq116$k, where each row of $\mathbf{H}$ indicates the channel samples captured periodically and columns represent the sample-value of each transmit antenna. 
\subsection{Performance Comparison of Conventional Vs. Hybrid Approach}
In this subsection, we compare the conventional versus the hybrid approach by using various evaluation parameters, which are given below.        
  
    
\subsubsection{NMSE}
To observe the effectiveness of the hybrid approach, we use NMSE, denoted by $\Omega_{\texttt{nmse}}$, which is the difference between the true channel and the acquired channel at the gNB. Mathematically,  
\begin{equation}
    \Omega_{\texttt{nmse}}=\mathsf{E}\left\{\frac{{} 	\left\|{{{{\mathbf{H}}}}-{\widehat{\mathbf{H}}_\text{gNB}}}\right\|^2_\text{FRO}}{{	\left\|{{{{\mathbf{H}}}}}\right\|^2_\text{FRO}}}\right\}\:
    \label{NMSE}
\end{equation}
where ${\widehat{\mathbf{H}}_\text{gNB}}$ denotes acquired channel at the gNB. Importantly, in the case of hybrid approach, acquired channel at the gNB is the one given in Equation\,\eqref{final}, whereas in the case of conventional approach, acquired channel at the gNB is compressed estimated channel, i.e., Equation\,\eqref{c_P_final}. 

\subsubsection{Precoding Gain}
As the feedback CSI is used for precoding, to observe precoding gain, we  use a simple matched filter precoder. Let us denote the acquired or reconstructed complex channel vector at the gNB as $\widehat{\mathbf{h}}_\text{gNB}$ and true complex channel vector as $\mathbf{h}$. By using this information, equivalent channel at the UE side can be represented as 
\begin{equation}
    \mathbf{h}_\text{eq}=\left(\frac{\widehat{\mathbf{h}}_\text{gNB}}{\left\|{{{{\widehat{\mathbf{h}}_\text{gNB}}}}}\right\|_{2}}\right)^{*}\times \left(\frac{\mathbf{h}}{\left\|{{\mathbf{h}}}\right\|_{2}} \right)
\end{equation}
where $\left\|{\cdot}\right\|_{2}$ denotes the Euclidean norm. Finally, precoding gain is calculated by
\begin{equation}
    \Gamma= \mathsf{E}\left\{(\mathbf{h}_\text{eq})^{*}\times \mathbf{h}_\text{eq}\right\}\:.
\end{equation}


\subsubsection{Spectral Efficiency}
Spectral efficiency is another important evaluation parameter to access the performance of precoding. The spectral efficiency, denoted by $\eta$, of a single-cell SU system is given by
\begin{equation}
    \eta=\log_{2}\left(1+\frac{\left|\mathbf{h}^{*}\cdot \mathbf{w}_{b}\right|^2}{\sigma^2}\right)
    \label{spectral_efficiency}
\end{equation}
where $\mathbf{w}_{b}$ is the beamforming vector, and $|\cdot|$ denotes the absolute value. As a symbol is multiplied by using a scalar digital precoder $p_\text{d}$ and an analog precoder $\mathbf{p}_\text{v}= \frac{\widehat{\mathbf{h}}_\text{gNB}}{\left\|\widehat{\mathbf{h}}_\text{gNB}\right\|_{2}}$, resultant $\mathbf{w}_b$ can be written as
\begin{equation}
    \mathbf{w}_b= {p}_\text{d}\cdot \mathbf{p}_\text{v}
\end{equation}
where the optimal value of ${p}_\text{d}=\sqrt{\frac{P}{N_t}}$ if there is a constraint of maximum transmit power ($|[\mathbf{w}_b]|^2\leq P$) and $|[\mathbf{p}_\text{v}]_{n_t}|^2=1$ \cite{Spectral_efficiency}. Therefore, Equation\,\eqref{spectral_efficiency} can be rewritten as
\begin{equation} 
\begin{split}
    \eta &=\mathsf{E}\left\{ \log_{2}\left(1+\frac{\left|\mathbf{h}^{*}\cdot \mathbf{p}_\text{v}\right|^2}{\sigma^2}\cdot\frac{P}{N_t}\right)\right\}\\
    &=\mathsf{E}\left\{\log_{2}\left(1+{\left|\mathbf{h}^{*}\cdot \mathbf{p}_\text{v}\right|^2}\cdot\frac{\gamma}{N_t}\right)\right\}\:
\end{split}
\end{equation}
where $\gamma=\frac{P}{\sigma^2}$ is SNR.
\subsubsection{Cosine Similarity}
To measure the quality of precoding vector, we consider the cosine similarity of the two vectors, which is calculated by 
\begin{equation}
    \rho=\cos{\theta}=\mathsf{E}\left\{\frac{|(\widehat{\mathbf{h}}_\text{gNB})^{*}\cdot \mathbf{h}| }{{\left\|{{{{\widehat{\mathbf{h}}_\text{gNB}}}}}\right\|_{2}}  {\mid\mid{{{\mathbf{h}}}}\mid\mid_{2}}}\right\}\:.
\end{equation}
\begin{figure}
\centering
\includegraphics[scale=0.39]{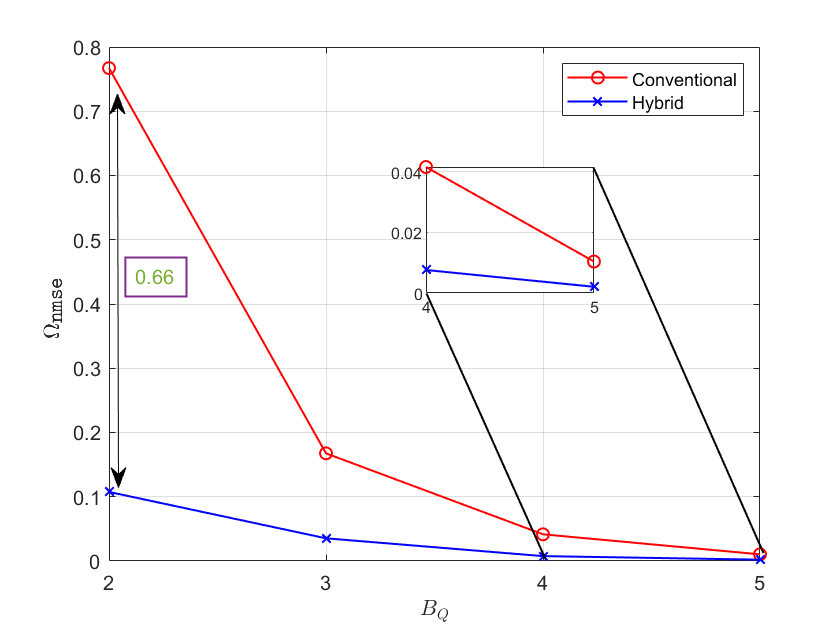}
\caption{A comparison of conventional CSI feedback mechanism vs. the hybrid approach with various quantization bits. For the sake of simplicity, here the performance is shown only for track-$1$.}
\label{Track-1_NMSE}
\end{figure}

In Fig.\,\ref{Track-1_NMSE}, benefits of the hybrid approach over the conventional approach can be observed, where $\Omega_{\texttt{nmse}}$ is plotted against different feedback bits (or say compression levels). The twofold advantages of the hybrid approach can be seen. The first advantage is when the compression is high, i.e., $B_Q=\,2$. It can be seen that the hybrid approach provides a gain of approximately $0.66$. However, a smaller gain is observed when the feedback overhead is increased. In short, using a massive number of feedback bits bring no advantage, which verifies the remark that $Q_c(x)=Q_h(x)=x$. The feedback followed in 3GPP is highly compressed; therefore, the hybrid approach is beneficial. On the other side, when $B_Q=\,5$, i.e., low compression, both methods show approximately zero error. But the hybrid approach brings the advantage of eliminating feedback, which is because the same CSI is already available at the gNB with the aid of channel predictor. In a nutshell, the hybrid approach reduces feedback overhead and minimizes compression errors under high compression, and it eliminates feedback when compression is low.

\begin{figure}
\centering
\includegraphics[scale=0.39]{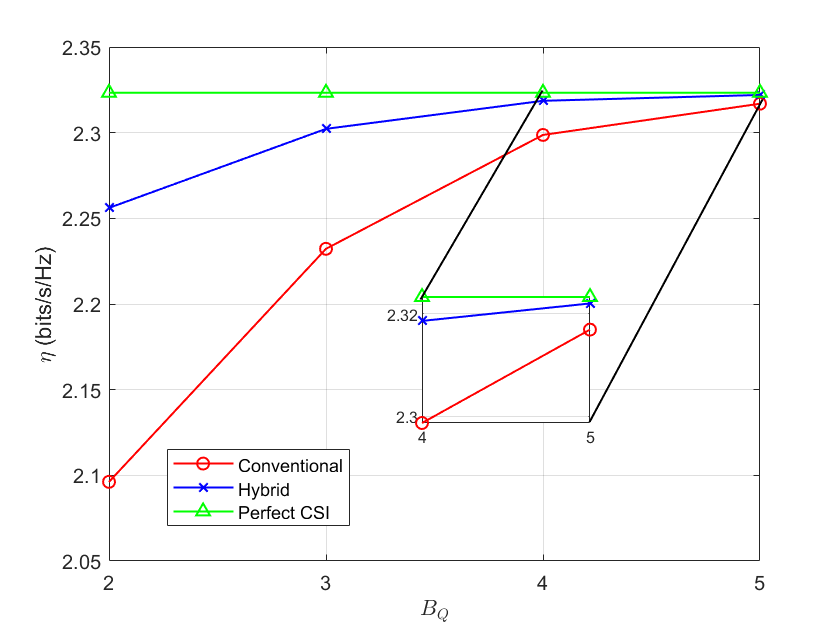}
\caption{Performance of spectral efficiency under different quantization bits on track-$1$. Here, maximum spectral efficiency that can be achieved is also plotted by assuming perfect CSI (i.e, no compression).}
\label{Track-1_SE}
\end{figure}

Fig.\,\ref{Track-1_SE} shows the performance of $\eta$ by using various feedback bits. It can be seen that the hybrid approach gives a higher gain in comparison to the conventional approach when compression is high, i.e., $B_Q =\,2$. In addition to this, the hybrid approach achieves maximum spectral efficiency by using four feedback bits, whereas the conventional approach reaches the goal by using one extra feedback bit. However, the hybrid approach eliminates feedback, while UE has to feedback in the conventional approach. In summary, the hybrid approach yields more gain under high compression and eliminates feedback requirements under low compression.   

\begin{figure}
\centering
\includegraphics[scale=0.39]{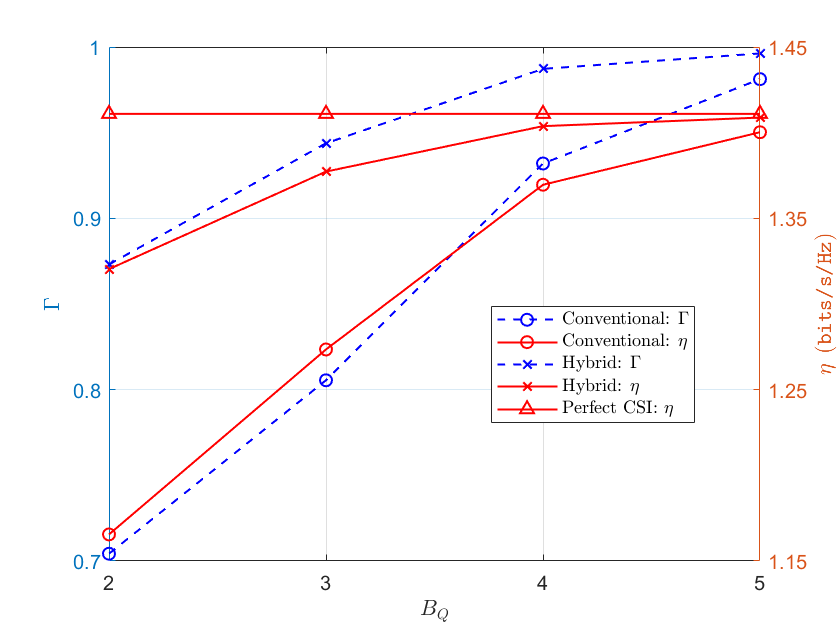}
\caption{A comparison of spectral efficiency and precoding gain vs. different number of quantization bits on track-$2$.}
\label{SE_PG}
\end{figure}

The performance of spectral efficiency, $\eta$, and precoding gain, $\Gamma$, can be seen in Fig.\,\ref{SE_PG}, where $\eta$ and $\Gamma$ are plotted on the right and left y-axis, respectively. The comparison of the conventional and hybrid approaches is shown by varying feedback bits. The increasing trend of $\eta$ and $\Gamma$ show that the performance gain is dependent on compression levels. However, the hybrid approach yields gain when feedback is highly compressed. For instance, when two quantization bits are used, then $\eta$ and $\Gamma$ provide a gain of approximately $0.16$ and $0.17$, respectively, compared to the conventional approach. The conventional and hybrid approaches' performance is almost identical when a high amount of feedback bits, i.e., $5$, are used or, say, low compression.

\begin{figure}
\centering
\includegraphics[scale=0.39]{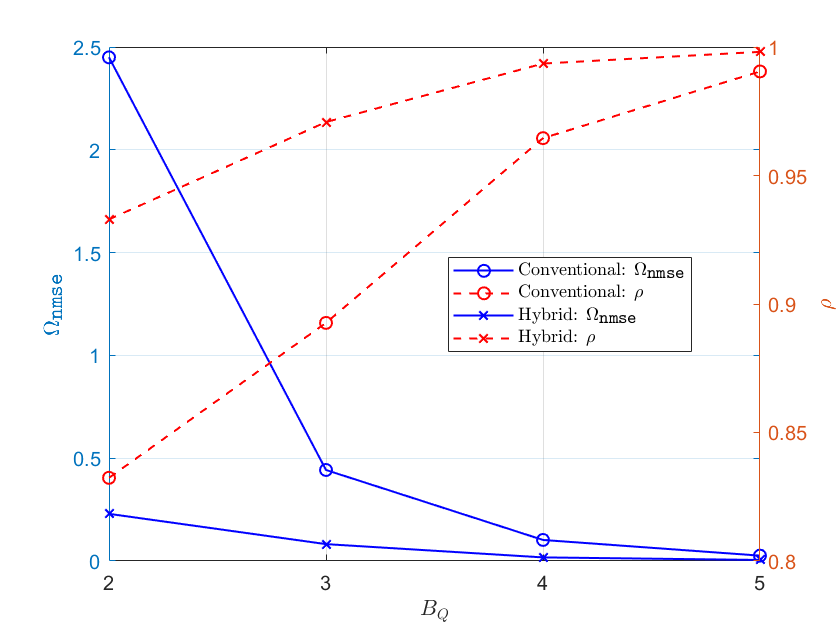}
\caption{Performance of NMSE and cosine similarity vs. number of quantization bits on track-$2$.}
\label{CS_NMSE}
\end{figure}

Fig.\,\ref{CS_NMSE} reveals the performance of NMSE and cosine similarity, depicted by $\Omega_{\texttt{nmse}}$ and $\rho$, respectively, on track-$2$. The trend of $\Omega_{\texttt{nmse}}$, plotted on the left y-axis, shows that the hybrid approach provides a huge gain of approximately $2.22$ under high compression, i.e., $B_Q=\,2$. And this gain reduces with the increase of overhead. Thus, the hybrid approach is highly beneficial when feedback information is massively compressed. In contrast, a similar gain can be observed for cosine similarity. For instance, when $B_Q=\,4$, then the hybrid approach has a cosine similarity of nearly $1$, which means that the acquired channel is parallel to the true channel, whereas the conventional approach reaches the hybrid approach by using one extra bit. In conclusion, trends of $\Omega_{\texttt{nmse}}$ and $\rho$ verify that the performance gain is high under high compression.


\begin{figure}
\centering
\includegraphics[scale=0.39]{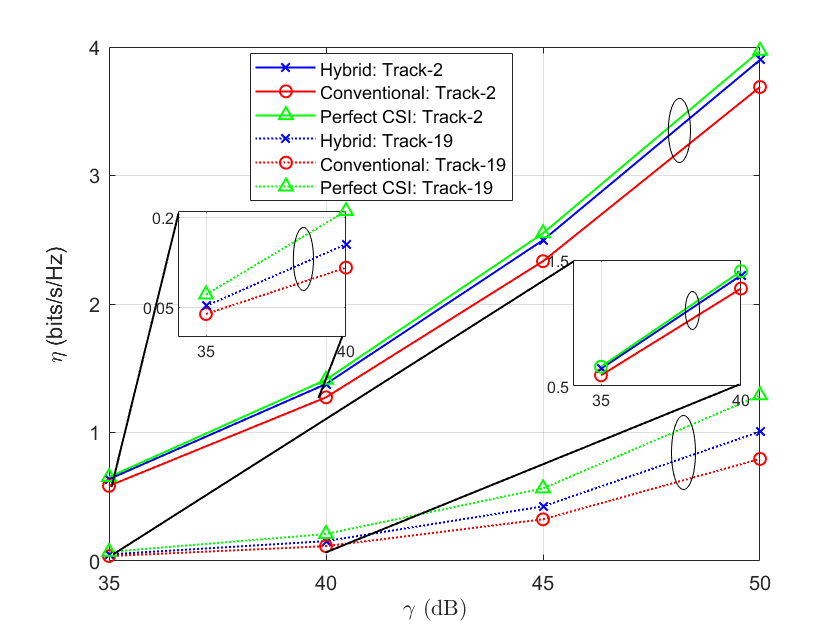}
\caption{Spectral efficiency vs. $\gamma$ on track-$2$ and $19$; solid-lines are representing track-$2$ whereas dotted-lines are depicting track-$19$. $B_Q=\,3$.}
\label{SE_SNR}
\end{figure}

In Fig.\,\ref{SE_SNR}, the performance of the conventional and the hybrid approach is compared using $\eta$ versus the SNR, $\gamma$, and by using the data-set of track-$2$ and track-$19$. The increase in spectral efficiency can be observed when $\gamma$ increases. Also, it can be observed that the hybrid approach's performance is almost similar to evaluated spectral efficiency under perfect CSI on track-$2$, whereas the hybrid approach lags on track-$19$, but it is still better than the conventional approach. In other words, the hybrid approach has recovered the true channel by nearly 100\% on track-$2$, while it is slightly worst on track-$19$. The reason for yielding lower gain on track-$19$ is due to not having good prediction as the predictor was trained on track-$1$, which is far from track-$19$. Besides, on track-$19$, we can observe smaller values of $\eta$ because the observed channel coefficients are smaller on track-$19$.  

In Table\,\ref{table:II}, the results of all evaluation parameters, i.e., $\Omega_\texttt{nmse}$, $\Gamma$, $\rho$, and $\eta$, are shown with various compression levels. The best results are presented in bold font, and the test data-set of track-$2$ is used, where maximum spectral efficiency that can be achieved under perfect CSI, i.e., no compression, is $1.411\,$(bits/s/Hz). It can be observed that the hybrid approach gives the lowest $\Omega_\texttt{nmse}$ values and outperforms the conventional approach. Similarly, the hybrid approach gives higher $\Gamma$ values for various compression levels. 
In addition, in the hybrid approach, cosine similarity is also maximum for all compression levels. Finally, the best values of $\eta$ also confirm the validity of the hybrid approach. For instance, when $B_Q=\,2$, the hybrid approach has a gain of approximately $0.16$, which decreases with the decrease in compression. In a nutshell, the hybrid approach outperforms the conventional approach. 

\begin{table}
\begin{center}
\caption{Performance comparison using four evaluation parameters}
\begin{tabular}{ c c|c c c c} 
\hline
$Q_B$ & Method & $\Omega_\texttt{nmse}\,\text{(dB)}$&$\Gamma$ & $\rho$&$\eta\,(\text{bits/s/Hz})$ \\
\hline


\hline
\multirow{2}{*}{2} & Conventional& $3.891$&$0.704$&$0.832$& $1.165$\\ 

&Hybrid&	$\mathbf{-6.394}$&$\mathbf{0.873}$&$\mathbf{0.932}$&$\mathbf{1.320}$\\ 
\cline{2-5} 
\hline
\multirow{2}{*}{3} & Conventional& $-3.539$&$0.805$&$0.892$&$1.273$ \\ 

&Hybrid&	$\mathbf{-10.856}$&$\mathbf{0.943}$&$\mathbf{0.970}$&$\mathbf{1.377}$\\ 


\hline
\multirow{2}{*}{4} & Conventional& $-9.896$&$0.932$&$0.964$&$1.369$ \\ 

&Hybrid&	$\mathbf{-17.5624}$&$\mathbf{0.987}$&$\mathbf{0.993}$& $\mathbf{1.403}$ \\ 


\hline
\multirow{2}{*}{5} & Conventional& $-15.8407$&$0.981$&$0.990$&$1.400$ \\ 

&Hybrid&	$\mathbf{-22.965}$&$\mathbf{0.996}$&$\mathbf{0.998}$&$\boldsymbol{1.408}$\\ 

\hline
\end{tabular}

\end{center}
\label{table:II}
\end{table}
\subsection{Computational Complexity}
To measure the complexity of the hybrid approach, we calculate the number of complex multiplications required for the training and prediction of channel predictor. To get a single-step prediction of the mMIMO channel, the hidden layer has to perform $I\times J$ complex multiplications, where $I$ and $J$ are the total number of neurons of the input and the hidden layer, respectively. We used two hidden layers; thereby, both hidden layers have to perform $J\times J$ complex multiplications. The complex multiplications at the output layer are $J\times K$, where $K$ are the total output neurons. Hence, the total complex multiplication are
\begin{equation}
    X=J\times(I+J+K)\:.
    \label{input_neurons}
\end{equation}

As the total number of input neurons, $I$, are dependent on the number of mMIMO sub-channels, recurrent components, and the delayed channel matrix. Hence, $I$ is calculated as

\begin{equation}
 \begin{aligned}
    I&=\underbrace{N_r\cdot N_t}_\text{ $\widehat{\mathbf{H}}^Q(t)$}+d\cdot\underbrace{(N_r\cdot N_t)}_\text{ $\widehat{\mathbf{H}}^Q(t-d)$}+\underbrace{N_r\cdot N_t}_\text{$Z^{-1}$}\\
    &=(d+2)\cdot N_r\cdot N_t\:.
    \label{in_neuron}
    \end{aligned}
\end{equation}
Conversely, the output layer's neurons are equivalent to mMIMO sub-channels; therefore, $K=N_r\cdot N_t$. By substituting Equation\,\eqref{in_neuron} and $K=N_r\cdot N_t$ into Equation\,\eqref{input_neurons}, $X$ can be written in the simplified form as
\begin{equation}
    \begin{split}
    X&=J\times\left[(d+2)\cdot N_r\cdot N_t+J+N_r\cdot N_t\right]\\
    &=J\times\left[N_r\cdot N_t\left(d+3+\frac{J}{N_r\cdot N_t}\right)\right]\:.
    \end{split}
\end{equation}
It is important to mention that the number of hidden layers and its neurons are dependent on the optimization problem; they are called hyperparameters of a ML algorithm. Generally, the same number of $J$ is used in all hidden layers. 

Let us denote the size of a mMIMO configuration by $M=N_r\cdot N_t$ and the scale of the channel predictor by $C=2dJ$. Then the one-step prediction complexity of the channel predictor is $\mathcal{O}(MC)$. The training complexity depends on the number of training examples (samples), denoted by $N_s$, and the number of training epochs, represented by $N_e$. Therefore, the training complexity of the channel predictor is $\mathcal{O}(MCN_sN_e)$.
\section{Conclusions and Future Outlooks}\label{conclusion}
Motivated by the inaccurate CSI acquisition, this paper addressed an ML-based mMIMO CSI feedback mechanism. The performance of the proposed approach is verified through a measurement campaign performed at the Nokia campus. The proposed approach showed that it can reduce feedback overhead and compression errors. We have learned that in a low compression scenario, the proposed approach can eliminate the feedback, and in the high compression, it brings more gain in comparison with the conventional approach. The numerical results evaluated using various parameters corroborated the validity of the proposed approach. Furthermore, we addressed the possible standardization points for the practical deployment of the proposed approach. 

Future works include evaluating the performance of the proposed approach exploiting 3GPP CSI reporting schemes, i.e., type-I and type-II. 


\ifCLASSOPTIONcaptionsoff
  \newpage
\fi




\bibliographystyle{IEEEtran}
\bibliography{mybib}

\end{document}